\documentclass[aps, prper, reprint,groupedaddress]{revtex4-2}  
\usepackage{graphicx, tabularx, longtable, nicefrac, amsmath,enumerate, supertabular, eurosym}
\usepackage{microtype}
\usepackage{booktabs}
\usepackage[utf8]{inputenc} 
\usepackage{csquotes}
\usepackage{booktabs}
\usepackage[normalem]{ulem}

\usepackage{hyperref,xcolor}

\begin{document}

\title{Harnessing Large Language Models to Enhance Self-Regulated Learning via Formative Feedback}

\author{Steffen Steinert$^1$}
\author{Karina E. Avila$^2$}
\author{Stefan Ruzika$^2$}
\author{Jochen Kuhn$^1$}
\author{Stefan K{\"u}chemann$^1$}
\email{mailto: s.kuechemann@lmu.de}

\affiliation{$^1$ Chair of Physics Education, Faculty of Physics, Ludwig-Maximilians-Universit{\"a}t M{\"u}nchen (LMU Munich), Geschwister-Scholl-Platz 1, 80539 Munich, Germany}
\affiliation{$^2$ Department of Mathematics, RPTU Kaiserslautern-Landau, Paul-Ehrlich-Str. 14, 67663 Kaiserslautern, Germany}

\date{\today}

\begin{abstract}Effectively supporting students in mastering all facets of self-regulated learning is a central aim of teachers and educational researchers. Prior research could demonstrate that formative feedback is an effective way to support students during self-regulated learning (SRL).  However, for formative feedback to be effective, it needs to be tailored to the learners, requiring information about their learning progress.
In this work, we introduce LEAP, a novel platform that utilizes advanced large language models (LLMs), such as ChatGPT, to provide formative feedback to students. 
LEAP empowers teachers with the ability to effectively pre-prompt and assign tasks to the LLM, thereby stimulating students’ cognitive and metacognitive processes and promoting self-regulated learning. We demonstrate that a systematic prompt design based on theoretical principles can provide a wide range of types of scaffolds to students, including sense-making, elaboration, self-explanation, partial task-solution scaffolds, as well as metacognitive and motivational scaffolds. In this way, we emphasize the critical importance of synchronizing educational technological advances with empirical research and theoretical frameworks.
\end{abstract}


\maketitle

\section{Introduction}
The study of formative feedback, a pedagogical approach that facilitates the bridging of gaps in understanding and fosters metacognitive processes in students, has a rich historical context and theoretical framework~\cite{butler1995feedback,narciss2008feedback,hattie2007power,wisniewski2020power}. In the contemporary era, the advent of Large Language Models (LLMs) presents a promising avenue for the implementation of effective intelligent tutoring systems that guide students through their learning journey. Despite their potential, LLMs have demonstrated certain biases~\cite{gallegos2023bias,jones2022capturing} and are prone to errors, particularly in the realms of physics and mathematics~\cite{kuchemann2023can,krupp2023unreflected}. 
A significant challenge is controllability and explainability. Developers and providers of LLM, as well as teachers who use them in classrooms, need to ensure that the output is accurate, safe and lawful. There are also concerns about students becoming overly reliant on the output without critically evaluating it.
Moreover, there are ethical considerations, such as concerns about bias and the impact of these models. For instance, racial biases, risks of misuse, inappropriate or unethical deployment, and loss of integrity are all valid concerns~\cite{kasneci2023chatgpt}.
Nevertheless, the continuous evolution and refinement of LLMs are gradually addressing these issues, bringing us closer to the reality of AI-powered tutors in educational settings~\cite{jeon2023large,hemachandran2022artificial}.


Such a development could revolutionize education by eliminating barriers such as geographical constraints, limited interaction, language barriers, and the fear of asking questions. It could also foster a more personalized learning experience. However, it is crucial to note that knowledge acquisition is not a straightforward process of information transfer~\cite{wittrock2000knowledge,everwijn1993ability}. 
Simply providing information via an interface like ChatGPT is insufficient, because the interaction with ChatGPT poses several demands on learners. These includes the design of a prompt and the critical evaluation of ChatGPT’s output. To truly facilitate learning, 
it’s essential to employ empirically validated theories and guidelines from fields such as pedagogy, didactics, and educational psychology.

Over the past years, self-regulated learning (SRL) 
has been the focus of intensive study in the field of educational psychology~\cite{lehmann2014cognitive,ifenthaler2012determining,zimmerman2002becoming}. One method to promote SRL is through the provision of formative feedback and various learning scaffolds, which are designed to stimulate cognitive processes and promote a deeper understanding~\cite{hattie2007power,wisniewski2020power}. In the context of personalized learning support, LLMs have demonstrated the capability to mimic different perspectives and roles through precise prompting~\cite{kieser2023educational}. This enables them to offer diverse types of feedback and scaffolding support to learners during their learning and problem-solving processes.

The rapid advancement of technology necessitates that educational research stays up-to-date of these developments to create effective educational tools. It is equally critical for researchers to persist in the development of open-source platforms in parallel with major corporations. This approach is essential to ensure the economic viability of these technologies, thereby preventing a digital disparity and ensuring equal opportunities for learners from a wide range of socio-economic backgrounds. In order to achieve this, it is important to harness these technologies in a manner that ensures their affordability and accessibility. 

In this paper, we introduce a platform called LEAP\footnote{\url{https://ai4physed.physik.uni-muenchen.de/LEAP}} (``Learning with AI about Physics") that enables teachers to generate content and interact with an LLM. Here, we demonstrate in detail the nuances in the prompts given the LLM so that it returns a specific type of formative feedback that facilitates different learning support mechanisms during self-regulated learning.
LEAP allows teachers to easily design problems in four short steps in which the LLM provides various forms of learning support, such as formative feedback, without the need for programming skills. The system can be prompted by teachers based on the type of cognitive process they wish to stimulate in their students. These prompts, which are invisible to the students, can be fine-tuned to provide various types of feedback. This approach also allows for a more controlled learning environment, reducing the potential for bias and hallucinations from the LLM. Furthermore, LEAP avoids the complication of instructing students to prompt ChatGPT, for instance in the openly available online chatbot interface, to provide specific types of learning support. 

While we currently use the application programming interface (API) of ChatGPT, the platform is designed to accommodate any LLM in the future. This flexibility allows users to select the model that best suits their needs. The API, essentially a set of rules that enable communication between different software applications, is a key component of this system.

At the time of writing, OpenAI has released new features for its ChatGPT Plus product offering the possibility to create a tutor with some of the capabilities of LEAP. This underscores the importance of educational research staying at the forefront of these technological advancements and investigate which type of learning support can be best implemented with LLMs. Our work contributes to this ongoing effort by providing a open-access tool that harnesses the power of most advanced opportunities of generative AI to enhance the learning experience.

\section{Theoretical Background}\label{SRL}
\subsection{Self-Regulated Learning and Formative Feedback}
{\em Self-regulated learning} (SRL) is a mental activity that enables individuals to control their thoughts, actions, and feelings to effectively steer their learning journeys~\cite{zimmerman2002becoming}. Possessing SRL skills equips a student with the ability to monitor their behavior in alignment with their goals and to self-reflect on their growing effectiveness. This self-monitoring and reflection are key to success, as they enhance the learners' ability to adapt and improve their strategies over time~\cite{zimmerman2002becoming}. To evolve into a self-regulated learner, an individual needs to cultivate the key processes: cognition, metacognition, and motivation~\cite{lehmann2014cognitive,ifenthaler2012determining,zimmerman2002becoming}. 


Cognition is the mental act of acquiring knowledge and understanding~\cite{ifenthaler2012determining,zimmerman2002becoming}. It encompasses processes like thinking, knowing, remembering, judging, and problem-solving. These are sophisticated functions of the brain that include language, imagination, perception, and planning.
Metacognition, in contrast, is the consciousness and comprehension of one's own cognitive processes~\cite{mahdavi2014overview}. It is a level above cognition and involves ``thinking about thinking"~\cite{mahdavi2014overview}. Metacognition empowers individuals to control their own learning by planning and monitoring their cognitive processes, checking progress, and adjusting strategies as necessary.
Motivation is the third element of SRL and is arguably its driving force~\cite{lehmann2014cognitive}. It is the eagerness or readiness to exert effort towards achieving objectives. Motivation in learning can originate from curiosity, interest in the subject matter, or the pleasure of mastering a new skill~\cite{mamlok2011can}.
These elements work together in a dynamic system to facilitate self-regulated learning. This process enables learners to take an active role in their learning, using strategies to acquire, comprehend, and remember information, as well as to evaluate and adjust their learning approaches as needed.


A crucial component to support SRL strategies among learners is {\em formative feedback}~\cite{nicol2006formative}. Formative feedback, which transpires during the learning process~\citep{narciss2008feedback}, is instrumental in enhancing learners’ understanding and refining their skills by stimulating learners’ metacognitive processes such as self-improvement and self-monitoring~\citep{butler1995feedback}. 
This feedback helps learners identify their strengths and weaknesses and target areas they need work on. It plays an important role in the learning process as it provides students with insights into their areas of proficiency and where additional effort is required~\citep{butler1995feedback}.

Kluger and colleagues highlight that formative feedback needs to be specifically related to the task or the learning process to be truly effective~\cite{kluger1996effects}. This involves providing information about the current level of performance or understanding, as well as guidance on how to close the gap between the present understanding and the targeted level of comprehension \cite{candel2020effects}. Formative feedback may also include different types of information such as cognitive (organizing elements of task requirements, elaboration of content), metacognitive (e.g., monitoring and self-regulation) or motivational aspects (e.g., relevance of addressing the task) \cite{candel2020effects}. 



\subsection{Cognitively activating activities}\label{CAA}

Cognitively activating activities are tasks that stimulate students’ cognitive processes~\cite{fortsch2017effects,hofer2018enhancing}. These processes include problem-solving, critical thinking, and understanding~\cite{gross2022cognitive}. The goal of these activities is to engage students’ minds in a way that promotes active learning and encourages them to think deeply about the material~\cite{fortsch2017effects}.

In physics, Hofer et al~\cite{hofer2018enhancing} have devised cognitively stimulating instructions to aid in the acquisition of a robust base of conceptual knowledge. They have developed five distinct types of instructions: generating solutions to novel problems, inventing with contrasting cases, comparing and contrasting, self-explanation prompts, and metacognitive questions. These methods have proven to be effective in enhancing the learning process.

Additionally, Smy and colleagues point out that the sense-making activities are relevant processes during the problem-solving and learning \cite{smy2016sensemaking}. These processes include problem defining, structuring, problem representation and generating potential alternate and the preferred solution. In this context, Berthold and colleagues also point out that organization and elaboration are two relevant cognitive strategies during learning \cite{berthold2007learning}. 


In essence, cognitively activating activities and formative feedback work together to create a dynamic and effective learning environment. They encourage students to be active participants in their learning journey, fostering a deeper understanding and mastery of the subject matter.

\subsection{Implementation of Principles of Self-Regulated Learning and Formative Feedback using LLMs into Learning Technologies}

When LLMs are implemented in learning platforms, they effectively can fulfill two important aspects. First, they can be effectively prompted to consider advanced learning principles, such as SRL and formative feedback, to enrich learning experiences and outcomes. In this way, LLMs have the potential to offer real-time formative feedback across all three dimensions of SRL - cognition, metacognition, and motivation - depending on the learners inputs and progress. This leads to a personalized learning experience that is tailored to learners' specific prior knowledge, competence levels, and attitudes. To foster SRL in students, 
strategic prompting can be employed to generate feedback in the form of cognitively activating instructions as suggested by Hofer and colleagues ~\cite{hofer2018enhancing}. For instance, by posing questions about students’ understanding, we can generate metacognitive scaffolds that encourage monitoring and self-reflection, as these content-specific scaffolds play a pivotal role in enhancing comprehension. Additionally, motivational prompts can be used to inspire and motivate students, thereby cultivating a conducive learning environment.

Second, in addition to personalizing the different levels of formative feedback, LLMs allow verbal interaction and stimulate learners to ask questions when they lack understanding or competencies. 

Combining both advantages, we introduce LEAP, a user-friendly interface designed to empower educators to create content and formulate prompts for ChatGPT. This enables the provision of specific types of formative feedback and enhances SRL. In this way, LEAP eliminates the need for programming skills, enabling educators to guide the feedback provided to students more effectively.


\section{Materials and Methods}
\subsection{Prompt design for LLMs based on theoretical principles}

\begin{figure}
    \centering
    \includegraphics[width=\linewidth]{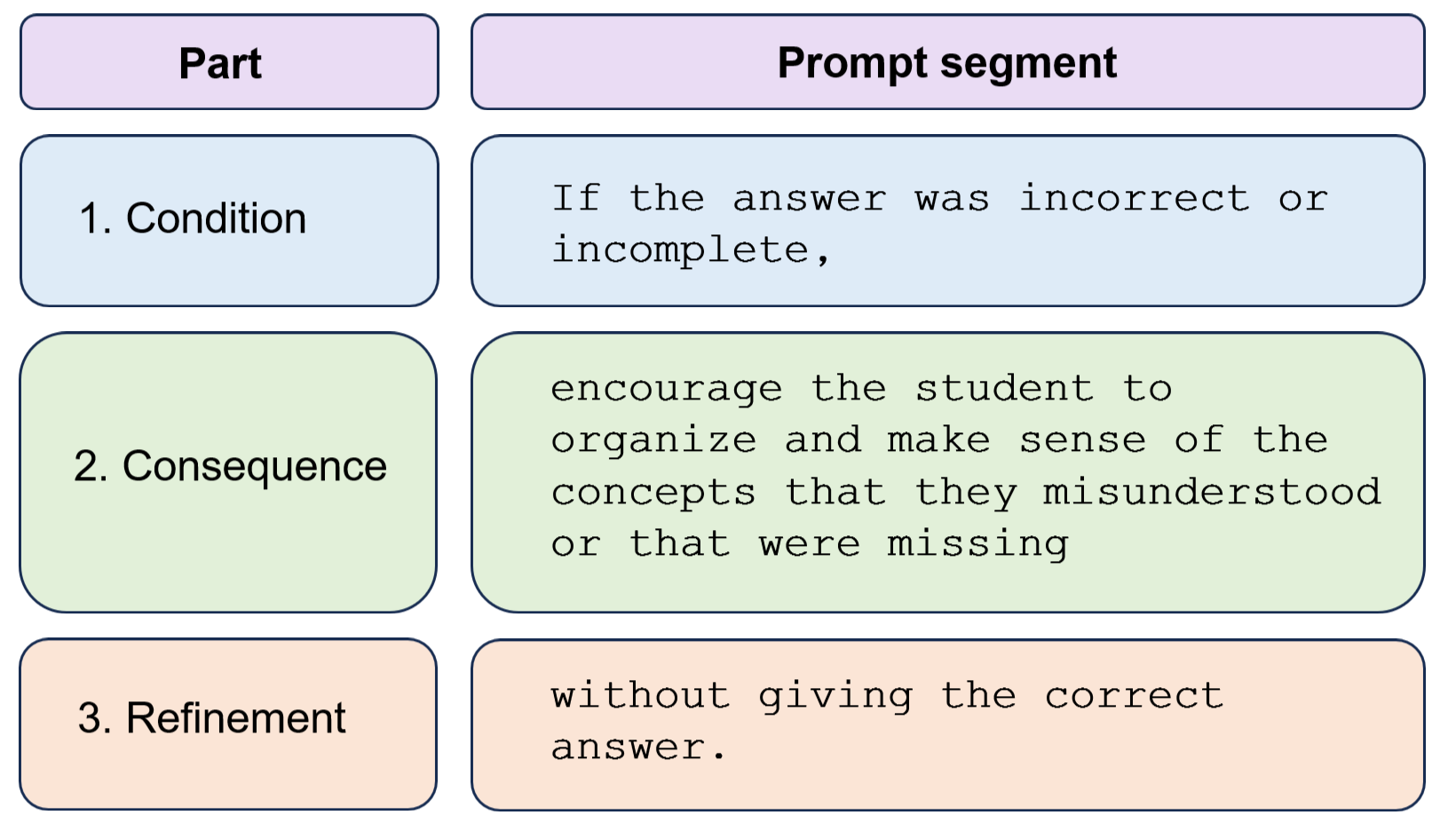}
    \caption{Systematic prompt generation in three parts based on theoretical principles.}
    \label{fig:prompt_generation}
\end{figure}

In this work, we systematically designed the prompts for the LLMs by splitting them up into three segments (Fig. \ref{fig:prompt_generation}) based on theoretical considerations and prior work. Here, a \textit{prompt} is defined as a sequence of instructions given to the LLM to shape the output (here: the formative feedback) provided by the LLM to the students. Each prompt begins with a \textit{Condition} that specifies the circumstances under which the LLM should  provide learning support to the student. Here, we used a statement such as ``If the students' answer is incomplete or incorrect". This segment ensures that learning support is only provided when necessary, as research has shown that adaptive learning support can be counterproductive if it merely reiterates the student’s existing knowledge, leading to an expertise reversal effect~\cite{kalyuga2009expertise}. This is because it consumes additional resources of the working memory leading to a decrease in learners' performance~\cite{kalyuga2009expertise}. 


The second part is the \textit{Consequence}, which outlines the expected outcome in case the condition is fulfilled. It comprises two elements: an adaptable one, chosen by the teacher or researcher, which requests a specific output (e.g., “encourage the student to …”), and an adaptive one, selected by the LLM, that is customized to the unique aspects of the student’s answer (e.g., “… the concepts that they misunderstood or that were missing”). In this way, the LLM is supposed to follow principles that are known from cognitive theory and prior educational research. 
The principles specifically incorporated here pertain to formative feedback and SRL (see Section~\ref{SRL}). These principles include strategies such as sense-making, elaboration, self-explanation, partial task solution, and the use of metacognitive and motivational scaffolds.  These strategies are usually employed as part of cognitively activating activities (see Section~\ref{CAA}).

The third part consists of a \textit{Refinement}, which includes aspects that are known to compromise the learning effectiveness of the considered theoretical principles in the second part. Here, we included the restriction ``without giving the correct answer". This restriction is motivated by prior research indicating that providing elaborated feedback in combination with the correct answer leads to a low efficiency of the feedback~\cite{kulhavy1985feedback,phye1989feedback} and may lead to superficial learning~\cite{schimmel1988patterns}.   

This approach to prompt design allows a systematic manipulation and empirical investigation of adaptive and adaptable segments of the prompt for the LLM. It ensures the LLM receives adequate information while eliminating superfluous elements. 

\subsection{Functions and Features of LEAP}\label{Functions}

To utilize the LEAP application, teachers are required to create an account by providing an email address and creating a password. Upon successful registration, users gain the ability to generate content. At the moment, users are free to choose between the ChatGPT version 3.5-Turbo and, in case they enter their personal OpenAI key, ChatGPT4.0-Turbo. Later, various LLMs with different fine-tunings will be provided to the teachers to choose from.

\begin{figure*}
    \centering
    \includegraphics[width=1\linewidth]{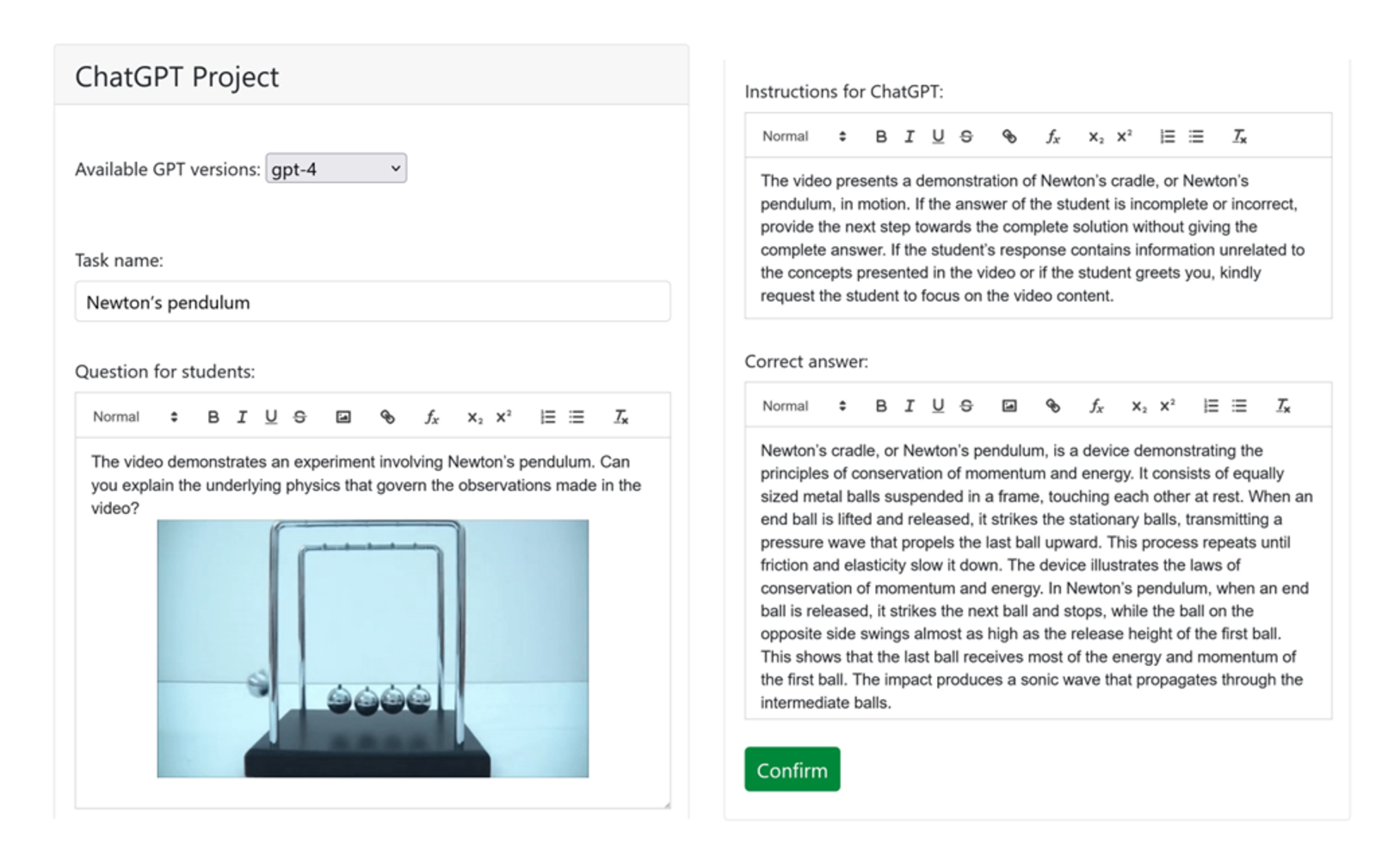}
    \caption{The LEAP window displays an example of how to assign a task to students. This process includes several steps: selecting the GPT version, naming the task, formulating the question for students, setting instructions for ChatGPT, and providing the correct answer.}
    \label{fig:leap_1}
\end{figure*}

To create a task with formative feedback or other educational content, teachers need to complete the following four fields (see Fig.~\ref{fig:leap_1}):
\begin{enumerate}
    \item Task Name: This refers to the title of the problem that the user intends to create.

    \item Question for Students: In this section, teachers can incorporate the question text, potentially accompanied by images, videos, mathematical equations or internet links. 

    \item Instructions for the LLM: This section allows teachers to specify the prompt type to support different mechanisms of self-regulated learning, such as cognitive sense-making or metacognitive scaffolds (see Section~\ref{results} for details). Also, this field should include a description of the images and videos incorporated in the question, providing the LLM with an understanding of what the students are viewing.

    \item Correct answer: In this field, teachers provide the correct answer to the task. This could be a detailed explanation delineating what constitutes an acceptable response from the student.
\end{enumerate}

Upon the creation of the content, the teacher can share a link with the students, ensuring the anonymity of the students is maintained. However, the teacher has the ability to review and download the history of the answers and the feedback provided by the LLM, as well as the history of the prompts used for the specific feedback.

When students click on the provided link, they are directed to a window showcasing the physics task created by the teacher. An example of such a task is depicted in Fig.~\ref{fig:leap}. The ‘Answer’ box allows students to submit their responses, which are then sent to the LLM to receive formative feedback. Students have the flexibility to interact with LLM as many times as they wish. Furthermore, they can choose the complexity level of the feedback (as explained in Section~\ref{implementation}) for each submitted response, based on their individual preferences.

\subsection{Architecture of LEAP}
LEAP is a web-based application that employs Flask\footnote{\url{https://flask.palletsprojects.com}} for backend development and SQLAlchemy\footnote{\url{https://www.sqlalchemy.org/}} for database access. The user interface is enhanced with the integration of frontend technologies such as HTML, CSS, Jinja2\footnote{\url{https://jinja.palletsprojects.com/}}, and JavaScript, ensuring an engaging and interactive user experience.

Our application’s database is constructed using MariaDB\footnote{\url{https://mariadb.org/}}, a widely recognized open-source relational database management system, ensuring reliable data management. To serve the application on the web, we use Gunicorn\footnote{\url{{https://gunicorn.org/}}}, a WSGI HTTP server, ensuring efficient handling of client requests.

Additionally, we employ Nginx\footnote{\url{https://nginx.org/}} as a reverse proxy, directing client requests to the Gunicorn server effectively. This comprehensive architecture guarantees an optimized user experience by efficiently managing client requests.

\subsection{ChatGPT implementation}\label{implementation}
ChatGPT is integrated into our Flask application using the API provided by OpenAI, facilitating a seamless interaction between the application and the AI model.

The messages sent to ChatGPT are organized as a list of dictionaries (i.e., specific prompts). Each dictionary represents a message, which consists of two key-value pairs: ``role" and ``content". The ``role" key identifies the sender of the message, which can be ``system", ``assistant", or ``user". The ``system" role sets the assistant’s behavior at the start of the conversation, the ``assistant" role signifies the AI model (ChatGPT) interacting with the user, and the ``user" role denotes the person communicating with the assistant. The ``content" key holds the actual content of the message.

In LEAP, the messages are structured as follows: 
\begin{enumerate}
    \item A ``system" message is sent, which includes:
    \begin{itemize}
        \item A scenario description: ``You are a friendly physics teacher having a conversation with a student. You are on first-name terms."
        \item Instructions for ChatGPT: These are provided by the educator and are explained in Section~\ref{Functions}.
        \item Complexity: This is an option that students can select to determine the complexity of the answer from ChatGPT. There are three categories:
        \begin{itemize}
            \item ``Low" instructs ChatGPT to ``Write a very easy to understand feedback in five sentences for a secondary school student".
            \item ``Middle" instructs it to ``Write your feedback for a high school student".
            \item ``High" instructs it to ``Write your feedback very elaborately and use physical formulas in MathJax format: \text{\$\$ $<$Formel$>$ \$\$}". MathJax\footnote{\url{https://www.mathjax.org/}} is an open-source JavaScript display engine that renders LaTeX, MathML, and AsciiMath notation in all modern browsers.
        \end{itemize}
    \end{itemize}
    \item An ``assistant" message presents the question asked to the student.
    \item A ``user" message provides the student’s answer to the question.
    \item Another ``assistant" message provides the correct answer to the question and a placeholder for the correction of the student’s answer.
\end{enumerate}


\section{Results}\label{results}
\begin{figure*}
    \centering
    \includegraphics[width=0.7\linewidth]{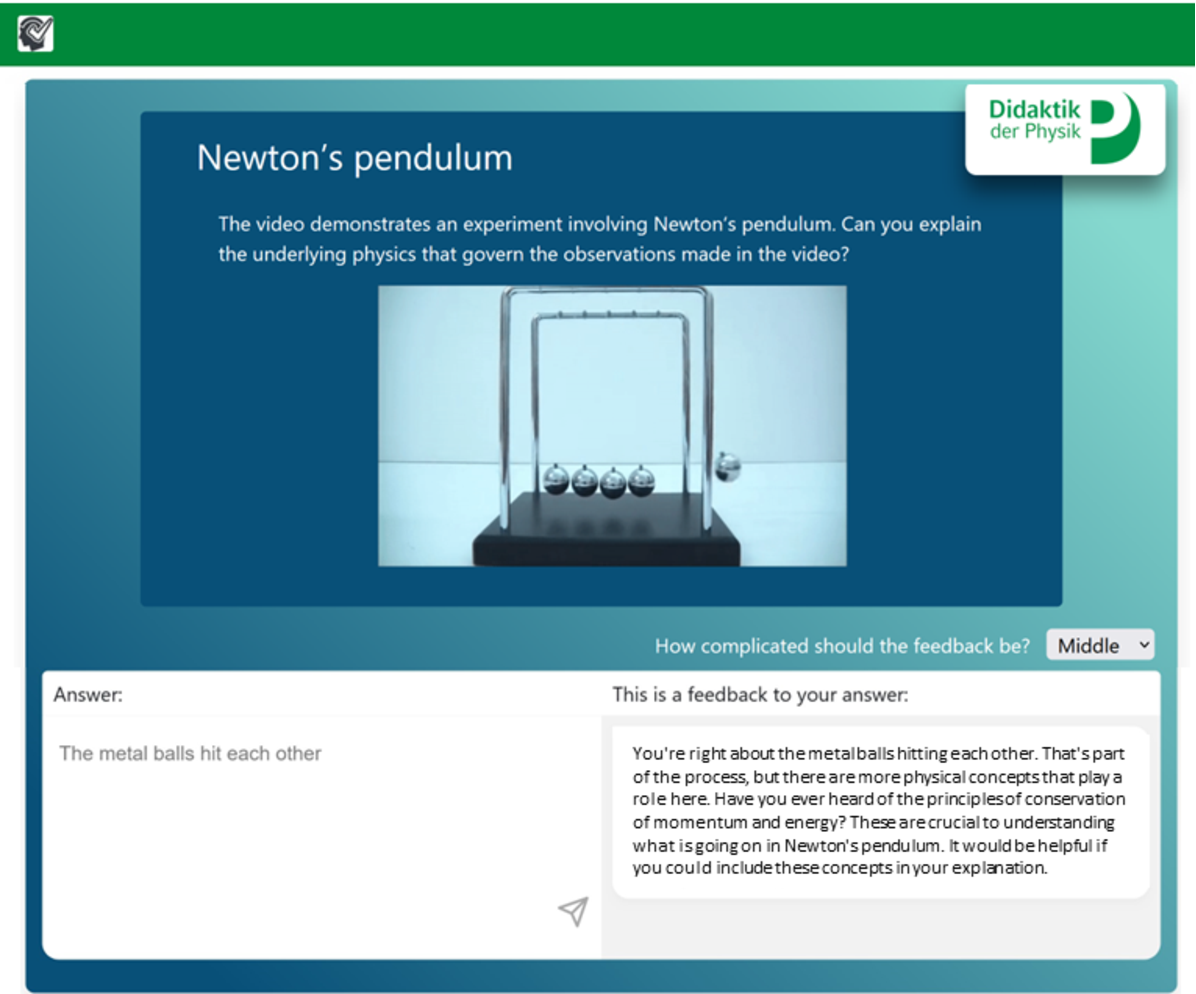}
    \caption{The LEAP window displays a physics task related to Newton’s cradle. Students can submit their responses in the ``Answer" box and send it to ChatGPT for feedback. The complexity of the feedback - ``low", ``middle", or ``high" - can be selected by the student based on their preference.}
    \label{fig:leap}
\end{figure*}

In this section, we examine the physics problem related to Newton’s cradle, as represented in Fig.~\ref{fig:leap}. In this particular task, students are shown a video that carries out an experiment involving Newton’s cradle. They are asked, ``The video demonstrates an experiment involving Newton’s cradle. Can you explain the underlying physics that govern the observations made in the video?"

Table~\ref{tab1} presents a series of prompt examples and the corresponding feedback generated by ChatGPT (version GPT-4 turbo) in response to the student's statement ``The metal balls hit each other". The prompts shown in this table correspond to the ``Instructions for ChatGPT" part presented in Section~\ref{Functions}. In addition to what is displayed in Table~\ref{tab1}, we instruct ChatGPT on the student’s visual context, for instance, ``The video presents a demonstration of Newton’s cradle, or Newton’s cradle, in motion". Subsequently, we provide the special prompt type shown in the table and finally instruct ChatGPT, ``If the student’s response contains information unrelated to the concepts presented in the video or if the student greets you, kindly request the student to focus on the video content". This final instruction is designed to prevent the generation of responses when the student inputs something unrelated to the task. 

The feedback provided by ChatGPT, as shown in Table~\ref{tab1}, demonstrate how strategic prompting can be formatted as cognitively activating instructions to potentially assist students in various self-regulating processes. By leveraging ChatGPT, students can receive cognitively activating instructions, including metacognitive, self-explanatory, or motivational feedback, as shown here. Moreover, these types of prompts can be combined to create a more tailored and effective feedback for the students. 

It is well-known that the quality of responses generated by ChatGPT is significantly influenced by the nature of the prompts provided. Within the LEAP interface, however, teachers are granted the flexibility to customize and test these prompts to optimize the AI’s responses. As expected, we found that more complex prompts tend to elicit superior results with advanced versions of ChatGPT.


\begin{table*}[ht!]
    \centering
    \caption{This table provides information on the type of prompts, the instructions given to ChatGPT, and the responses generated for the physics problem related to Newton’s Cradle. The feedback provided by ChatGPT is in relation to the answer statement ``The metal balls collide with each other".}\label{tab1}
    \begin{tabular}{l @{\hspace{.5cm}} p{6.1cm} @{\hspace{.5cm}} p{6.1cm} }
     \hline \hline \\[-0.1cm]

			\textbf{Theoretical aspect}	& \textbf{Prompt for ChatGPT}	&  \textbf{ChatGPT feedback}\\
	
                   \midrule

Cognitive -- Sense-making~\cite{smy2016sensemaking}	& If the student's answer is incomplete or incorrect, encourage the student to organize and make sense of the concepts that they misunderstood or that was missing without giving the correct answer. & You're right, the metal balls hit each other. But that is only one part of the whole. Think about what happens when the balls hit each other. What happens to the energy and momentum? And why does only the last ball move and not the others? Try to answer these questions to get a better understanding of Newton's cradle.\\

                     \midrule
                    
 Cognitive - Elaboration~\cite{berthold2007learning} 	&  If the student's answer is incomplete, encourage the student to elaborate on aspects that were missing without giving the correct answer. & You're right, the metal balls hit each other. But that's only part of the story. Can you perhaps say a little more about this? For example, what happens when one ball hits the others? And why does the last ball move even though it wasn't hit directly by the first ball? Think about the laws of physics that we discussed in class.\\
    \midrule
Cognitive - Self-explanation~\cite{hofer2018enhancing}	& If the answer of the student is incomplete or incorrect, give an example of a simpler situation in which the incorrect or missing concept plays a central role and ask the student how he would explain the situation. Do not give the correct answer. & You're right about the metal balls hitting each other. But what happens after that? Think of a row of dominoes. What happens when you knock over the first domino?\\
 
 \midrule
 
Cognitive - Partial task solution	& If the answer of the student is incomplete or incorrect, provide the next step towards the complete solution without giving the complete answer. & That's a good start! You have correctly recognized that the metal balls hit each other. But what exactly happens when they do? Think about the laws of physics that could play a role here. What can you say about the law of conservation of energy and momentum?\\
  
                \midrule

Metacognitive~\cite{hofer2018enhancing,berthold2007learning} & If the student's answer is incomplete or incorrect, ask specifically if the student understands the role of the concepts that they misunderstood or that were missing without giving the correct answer. & You're right about the metal balls hitting each other. That's part of the process, but there are more physical concepts that play a role here. Have you understood the principles of conservation of momentum and energy? These are crucial to understanding what is going on in Newton's cradle. It would be helpful if you could include these concepts in your explanation.\\

	                   \midrule
                    
Motivational~\cite{hawthorne2015effects}   & If the student's answer is incomplete or incorrect, motivate them by reminding them that every attempt to understand something is a step in their learning journey. Encourage them to keep going and stay curious. Do not give the correct answer or any hints.  & That's a good start! You have correctly recognized that the metal balls hit each other. Now try to delve deeper into the topic and explain what happens after the impact and why. Keep at it and stay curious!\\

   \\ \hline \hline
    \end{tabular}
   
\end{table*}

\section{Discussion}

The alignment of empirical research and theoretical models with technological advances is vital for enhancing learning experiences and ensuring the future relevance of educational systems. Following this approach, we developed LEAP, a platform to harness the great opportunities of LLMs for personalization and interaction to support students during self-regulated learning. Platforms such as LEAP act as a bridge for teachers, enabling them to effectively implement LLMs and facilitate the creation of tailored content for students. LEAP has the potential to provide real-time, personalized formative feedback, supporting various aspects of self-regulated learning, including cognition, metacognition, and motivation. We demonstrated that specific prompting allows the provision of these specific cognitively activating scaffolds based on theoretical principles. 


Through platforms like LEAP, LLMs could soon be incorporated into classroom activities. This integration would provide personalized formative feedback to students in a controlled environment, stimulating various aspects of self-regulated learning. For teachers, the incorporation of such platforms into their technological toolkit offers valuable resources to enhance their instructional practices. The utilization of this technology has the potential to free up teachers’ time, allowing them to focus on other critical aspects of teaching, while providing personalized targeted support. 

In comparison to the chatbot version of ChatGPT and other LLMs, LEAP does not require students to learn about prompting before using LLMs for learning. The specific pre-prompting overtakes various time-consuming but relevant tasks of teachers. At the same time, it strongly reduces the number of erroneous output and hallucinations because the teacher provides the correct solution to the task, and it does not allow the student to get side tracked as it kindly guides the student back to the task. It has also the advantage that learners do not need to critically reflect as in a chatbot as the output is aligned with the correct solution. 

LEAP can also support academic research, as it enables teachers and researchers to extract data for research purposes. This does not only facilitate academic research but also offers teachers valuable insights into how students interact with the content. These insights can guide interventions or adjustments to the prompting strategy, thereby enhancing the effectiveness of the learning process. This underscores the adaptability and potential of our system in both educational and research contexts.

Furthermore, with the advent of new technological advances, such as the release of GPT-4 Turbo, we can anticipate a future where LLMs become more accurate and provide an enhanced user experience. These advances will likely introduce new functionalities, paving the way for more sophisticated and effective AI tutoring systems. Here, we are free to incorporate any newly developed LLM into LEAP being able to have a platform that is easily to maintain up to date. As we continue to refine and develop these technologies, the potential for AI to revolutionize education becomes increasingly tangible.


At the moment, it is up to the teacher or researcher to decide which type of formative feedback LEAP should provide. However, it remains a matter of debate whether a single type of formative feedback would be equally beneficial for all learners.
It is rather likely that specific learner characteristics would influence the effectiveness of the feedback.

Finally, the affordability, usability and ease-of-use of such systems are critical considerations. While our application currently utilizes ChatGPT, LEAP will soon be adapted to incorporate LLMs from other providers, particularly free LLMs, thereby increasing accessibility. Although LEAP is currently accessible to everyone and allows free use of ChatGPT 3.5-Turbo within the platform, there are still ongoing expenses associated with using other version of OpenAI’s API. 
To ensure that everyone can benefit from the ongoing technological revolution with LLMs in education, it is necessary to extent the options of LLMs within LEAP to include more free versions. In this regard, it could be beneficial to include LLMs that have been specifically adapted or ``fine-tuned" to accommodate certain fields of study or areas of interest.

\section{Conclusion}
In this work, we introduced a platform named LEAP that utilizes LLMs, currently ChatGPT, to provide feedback that supports students in various aspects of SRL, including cognition, metacognition, and motivation. This platform simplifies the interaction between teachers and LLMs, enabling them to easily create tasks that stimulate SRL in students via fine prompting. Our findings suggest that the integration of LLMs into education is a promising avenue with the potential to revolutionize teaching and learning practices. By making education more personalized, effective, and accessible, platforms like LEAP are paving the way for a new era in educational technology.

\section{References}


\bibliography{references.bib}

\end{document}